\documentclass[12pt]{article}
\usepackage[utf8]{inputenc}
\usepackage{graphicx} 
\usepackage{dcolumn} 
\usepackage{bm} 
\usepackage{amsfonts}
\usepackage{amsthm}
\usepackage{amsmath}
\usepackage{amssymb}
\usepackage{epsfig}
\usepackage{color}
\usepackage{textcomp}
\usepackage{hyperref}
\usepackage{titlesec}
\usepackage{slashed}
\usepackage{caption}
\usepackage{subcaption}
\usepackage{booktabs}
\usepackage{multirow}
\usepackage{cite}
\usepackage{placeins}
\usepackage{comment}
\usepackage{braket}
\usepackage{authblk}
\usepackage{physics}
\usepackage{mathtools}
\numberwithin{equation}{section}

\title{Infeasibility of Graviton Detection as Cosmic Censorship}
\author{Andrea Palessandro \thanks{andrea.palessandro@gmail.com}}
\affil{\small Società Italiana di Fisica}
\date{}

\begin{document}

\maketitle

\begin{abstract}
    \noindent We construct an explicit model of inhomogeneous gravitational collapse leading to a naked singularity in which gravitational absorption is both efficient and observable. We propose that the infeasibility of graviton detection is simply a consequence of Nature's conspiracy to hide regions of strong curvature behind event horizons.
\end{abstract}

\section{Introduction}
In the theory of General Relativity, gravitational singularities are both inevitable and ubiquitous \cite{Penrose, Hawking}. The Cosmic Censorship Conjecture (CCC) posits that these singularities are always hidden from view behind an event horizon\footnote{The technical statement is that for generic initial data, the maximal Cauchy development of a solution of Einstein’s equations possesses a complete future null infinity.} \cite{Penrose2, Penrose3}. Given that quantum gravitational effects are expected to become visible in regions of large space-time curvature, the CCC is sometimes taken to imply the unobservability of quantum gravity \cite{Joshi0}. If this view is correct, an observer could probe the singularity and establish the quantization of gravity, but would not be able to communicate their results with the outside world due to the existence of an uncrossable event horizon. 

Conceptually, the simplest experiment that could be performed to prove the quantum nature of gravity is the detection of single gravitons \cite{Dyson, Rothman, Carney, Tobar, Pal, Carney2}. This can be done, for example, by sending gravitational radiation through a cloud of atoms. If the gravitational field is quantized, gravitons of a certain wavelength will be absorbed, resulting in absorption lines in the gravitational spectrum \cite{Pal2}, a strong indicator of the field's granularity\footnote{Granular absorption by quantized matter can in principle take place in semi-classical gravity too \cite{Carney}. However, there are tests one could perform that would be able to distinguish between a purely classical model of gravitational radiation and a quantum one; for example, by observing sub-Poisson graviton counting statistics. The deviation from Poisson statistics is proportional to the efficiency of the detector\cite{Carney2}, so this would only work if the detector is $\sim \mathcal{O}(1)$ efficient, but this is exactly the condition needed to have visible absorption lines.}.

Given a cloud of atoms of constant density $\rho$ and total mass $M$, the optical depth of a graviton traveling through the cloud is \cite{Dyson}
\begin{equation}\label{tau}
    \tau = n \sigma R,
\end{equation}
where $n \equiv \rho/\mu$ is the number density of atoms, $\mu$ their mass, $\sigma \sim G$ the gravitational absorption cross section, and $R$ the extension of the cloud. Assuming the Compton wavelength of a single atom is contained within the cloud, $\mu R > 1$, imposing $\tau > 1$ gives $R < GM$, which is the condition for gravitational collapse. This means that the parameter space that allows for graviton detection via visible absorption lines ($\tau > 1$) corresponds to a collapsed atomic cloud and is thus usually considered to be hidden from view\footnote{Note that, contrary to what is commonly assumed, the difficulties associated with graviton detection are \textit{not} due to gravity being weak. If gravity were stronger, it would be easier to absorb gravitons, but it would also be correspondingly easier to create black holes. The maximum number density before gravitational collapse ensues is $n_{B, \text{max}} \sim (\mu G R^2)^{-1}$, while the absorption cross section is $\sigma \sim G$. The two quantities scale in opposite ways with $G$, therefore the maximum optical depth is independent of the strength of the gravitational coupling.}.

The conclusion rests crucially on the assumption that gravitational collapse generically results in black holes, i.e. singularities hidden behind event horizons. Were this not true, for example in the case of a naked singularity, one could imagine an experiment in which graviton absorption is both efficient ($\tau > 1$) \textit{and} visible from the outside. The aim of this paper is to demonstrate this by explicit construction.

\begin{figure}[ht]
    \centering
    \includegraphics[width=0.9\linewidth]{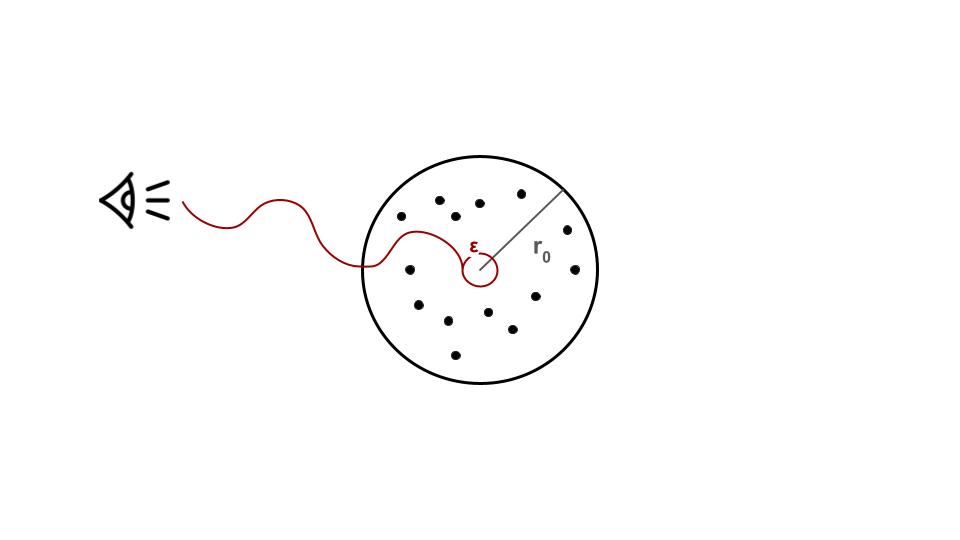}
    \caption{The experimental setup. An observer looks for absorption lines in the gravitational spectrum of radiation emitted by a source close to the center of an inhomogeneously distributed atomic gas cloud.}
    \label{fig1}
\end{figure}

The experiment's setup is described in Figure \ref{fig1}. The detector is a gas cloud of extension $r_0$ and total mass $M$ with an inhomogeneous mass distribution that slowly collapses under its own gravity. A gravitational wave source is placed at distance $r = \epsilon r_0 \ll r_0$ from the central singularity, emitting radiation to infinity. An observer placed outside the cloud analyzes the incoming gravitational radiation and looks for absorption lines to establish the quantization of gravity. Clearly, two conditions have to be satisfied in order for this experiment to be successful:
\begin{itemize}
    \item The detector has to be efficient, i.e. $\tau > 1$ for a graviton traveling through the gas cloud.
    \item The gravitational radiation ought to escape the gas cloud, meaning that the central singularity has to be (globally) visible.
\end{itemize}

In the rest of the paper we will construct an explicit example of inhomogeneous spherically symmetric gravitational collapse in which both conditions above are satisfied. In particular, in \S \ref{LTBspace} we study the evolution of the detector (gas cloud) in the special case of a spherically symmetric mass distribution. In \S \ref{local} we work out the necessary conditions for the local visibility of the central singularity, while in \S \ref{global} the conditions for its global visibility. In \S \ref{absorption} we demonstrate the efficiency of the detector if the gravitational source is placed sufficiently close to the (naked) singularity. Finally, we present our concluding remarks in \S \ref{conclusions}.

\section{Spherically symmetric gravitational collapse}\label{LTBspace}
In this section we study a specific model of spherically symmetric gravitational collapse which can give rise to a naked singularity \cite{Christodoulou, Joshi}. In the two sections that follow we will specify the conditions for this to happen both locally and globally. We work in natural units $\hbar = c = 1$ and with metric signature $(-,+,+,+)$. Dots and primes indicate differentiation with respect to time and space, respectively.

The general class of solutions describing the evolution of a spherically symmetric inhomogeneous dust cloud is given by the Lemaitre-Tolman-Bondi metric \cite{Lemaitre, Tolman, Bondi}
\begin{equation}\label{LTB}
    ds^2 = -dt^2 + \frac{R'(t,r)^2}{1-k(r)} dr^2 + R(t,r)^2(d \theta^2 +\sin^2\theta d\phi^2),
\end{equation}
where $R(t,r)$ is the proper radius of a matter shell at comoving coordinates $(t,r)$, and $k(r)<1$ controls the curvature of the spatial slices at constant $t$.

The metric is sourced by the energy-momentum tensor of a pressureless fluid:
\begin{equation}\label{Tmn}
    T_{\mu \nu} = \rho(t,r) \delta^0_\mu \delta^0_\nu,
\end{equation}
where $\rho(t,r)$ is the matter density of the dust cloud. We assume the cloud is made up of atoms, with total mass $M$.

The Einstein field equations give \cite{Enqvist}
\begin{equation}\label{integral}
    \frac{\dot{R}^2+k}{R^2} = \frac{2 G m}{R^3}.
\end{equation}
where $m(r)$ is the mass enclosed in a sphere of radius $R(r)$:
\begin{equation}\label{constraint}
    m' = 4 \pi R^2 R' \rho.
\end{equation}

The model is called bound, marginally bound or unbound depending on whether $k>0$, $k=0$ or $k<0$. For pedagogical clarity, we analyze here the marginally bound case. Integration of (\ref{integral}) with $k=0$ gives 
\begin{equation}\label{solution}
    R(t,r) = \left( r^{3/2} - \frac{3}{2} \sqrt{2Gm} \, t\right)^{2/3}.
\end{equation}
Note that, since we are interested in gravitational collapse, we have taken the solution with $\dot{R}<0$. Moreover, we have used the remaining coordinate freedom to equate proper and coordinate distance on the initial hypersurface, i.e. $R(0,r) = r$.

Clearly, the mass function is fixed once the initial density distribution $\rho(0,r) \equiv \rho(r)$ is given:
\begin{equation}
    m(r) = 4 \pi \int \rho(r) r^2 dr .
\end{equation}
We further assume that $\rho(r)$ is of the form
\begin{align}\label{rho}
\begin{split}
    \rho(r) &= \rho_0 \left[1-\left(\frac{r}{r_0}\right)^n\right] \quad \text{for} \quad 0 \leq r \leq r_0,\\
    \rho(r) &= 0 \quad \text{for} \quad r>r_0,
\end{split}
\end{align}
where $r_0$ is the initial extension of the gas cloud, and $\rho_0$ the initial matter density at $r=0$. By Birkhoff's theorem the spacetime at $r>r_0$ is described by the Schwarzschild metric with total mass $M$.
If $n>0$, the density of the cloud decreases monotonically as one moves out from the center. Given (\ref{rho}), the mass function is
\begin{equation}\label{mass}
    m(r) =  M \left[ 1+ \frac{3}{n} - \frac{3}{n} \left( \frac{r}{r_0}\right)^{n} \right] \left( \frac{r}{r_0}\right)^3,
\end{equation}
where $m(r_0) = M \equiv 4\pi n \rho_0 r_0^3/(3(n+3))$ is the total mass of the cloud. 

Gravitational singularities are defined as points at the boundary of spacetime where the energy density or the curvature scalars diverge. One such example is the Kretschmann scalar $\mathcal{K} = R_{abcd}R^{abcd}$, which for the metric (\ref{LTB}) is given by
\begin{equation}\label{kert}
\begin{split}
    \mathcal{K} &= 4\frac{(k+\dot{R}^2)^2}{R^4}+8\frac{\ddot{R}^2}{R^2}+2\frac{(k'+2\dot{R}\dot{R}')^2}{R^2R'^2} + 4 \frac{\ddot{R}'^2}{R'^2}\\
    &= 48 \frac{G^2 m^2}{R^6} - 32 \frac{G^2 m m'}{R^5R'} + 12 \frac{G^2 m'^2}{R^4 R'^2},
\end{split}
\end{equation}
where the second equality follows from (\ref{integral}). Clearly, the Kretschmann scalar diverges for both $R = 0$ (with $R', m' \neq 0$) and $R' = 0$ (with $R,m' \neq 0$). 

The former is called a shell-focusing singularity and takes place when the physical radius of a matter shell shrinks to zero. According to (\ref{solution}), this happens at the time
\begin{equation}\label{t_c}
    t_c(r) = \frac{2}{3} \frac{r^{3/2}}{\sqrt{2Gm}}.
\end{equation}
This is the time of collapse for a matter shell at comoving distance $r$ from the center. In general, for an inhomogeneous mass distribution different shells will meet the singularity at different times depending on the value of $r$.

The latter is called a shell-crossing singularity, and generically occurs whenever $t_c(r)$ is not a monotonically increasing function, i.e. when matter shells cross \cite{Newman, Joshi3}. At a crossing event the matter density and certain components of the Riemann curvature tensor blow up, but the causal structure of spacetime can be extended through it \cite{Nolan}. Unlike shell-crossing singularities, spacetime admits no extension through a shell-focusing singularity, which is therefore the only type of ``genuine'' singularity in a causal sense \cite{HawkingEllis}. Given the mass function (\ref{mass}), the time of collapse (\ref{t_c}) is a monotonically increasing function of $r$, therefore no shell-crossing singularities occur in our model. We focus then on the shell-focusing singularities.

In order to determine the nature of the singularity (hidden or naked), one must study the behavior of outgoing non-spacelike geodesics in the spacetime (\ref{LTB}). For simplicity we will focus on outgoing radial null geodesics, which, for $k=0$, are described by
\begin{equation}\label{geodesic}
    \frac{dt}{dr} = R'.
\end{equation}
Along an outgoing radial null geodesics during the collapsing phase we have \cite{Harada}
\begin{equation}
    \frac{dR}{dt} = \dot{R} + R' \frac{dr}{dt} = 1 - \sqrt{\frac{2Gm}{R}}.
\end{equation}
Therefore, $dR/dt < 0$ whenever $R<2Gm$ and the corresponding region is trapped, meaning that all light rays converge towards the singularity. The outer boundary of the trapped region is called the apparent horizon, and lies at $R=2Gm$. Given (\ref{solution}), the apparent horizon forms at time
\begin{equation}
    t_{ah}(r) = t_c(r) - \frac{4}{3} G m.
\end{equation}
Clearly then, $t_{ah}(r) \leq t_c(r)$ in any neighborhood of $r \neq 0$, therefore a non-central shell-focusing singularity is always hidden. A central shell-focusing singularity can be locally naked if in a neighborhood of $r = 0$, $t_{ah}(r) > t_c(0)$, so that an outgoing radial null geodesics can probe the singularity without encountering any trapped surface. Similarly, the singularity can be globally naked (visible to observers at infinity) if the validity of the condition $t_{ah}(r) > t_c(0)$ extends to the outer edge of the gas cloud. This condition, however, only provides a necessary (but not sufficient) criterion for visibility, as we will explain in the next section.

\section{Local Visibility}\label{local}
As we discussed in the previous section, only the central shell-focusing singularity can be naked. This forms at the time
\begin{equation}\label{t0}
    t_c(0) \equiv t_0 = \frac{2}{3} \sqrt{\frac{n}{n+3} \frac{r_0}{2GM}}r_0 =\frac{1}{\sqrt{6 \pi G \rho_0}}.
\end{equation}
Near $r=0$, we can write (\ref{t_c}) as
\begin{equation}\label{tc}
    t_c(r) = t_0\left[1 + \frac{3}{2(n+3)} \left( \frac{r}{r_0}\right)^n \right] + \mathcal{O}(r^{n+1}).
\end{equation}
Since $t_c(r) \geq t_0$, the central singularity at $r=0$ forms first, followed by the outer shells, in order of distance from the center. The limiting case $n \rightarrow \infty$ corresponds to homogeneous collapse (the Oppenheimer–Snyder model \cite{Oppenheimer}) with $\rho(r) = \rho_0$, in which all matter shells collapse simultaneously at $t_c(r) = t_0$ regardless of $r$. 

In order for the central singularity to be naked, at least locally, it is necessary for the apparent horizon to form \textit{after} collapse, i.e. $t_{ah}(r) > t_0$ in a neighborhood of $r=0$. Near $r=0$, the apparent horizon forms at the time
\begin{equation}
    t_{ah}(r) = t_c(r) - \frac{4}{3} G m = t_0 - \frac{4(n+3)}{3n} GM \left(\frac{r}{r_0}\right)^3  + \frac{3 t_0}{2(n+3)} \left( \frac{r}{r_0}\right)^n +\mathcal{O}(r^{n+1}).
\end{equation}
For $n \rightarrow \infty$ (homogeneous collapse) $t_{ah}(r) \leq t_0$ and the singularity is hidden. For finite values of $n$, if $n=1,2$, $t_{ah}(r) \geq t_0$ around $r=0$ and the singularity is potentially naked. If $n=3$ the singularity is potentially naked only when $t_0 > 32/3 \, GM$, or $r_0 \gtrsim 10 \, G M$. For $n \geq 4$ the singularity is again hidden behind an event horizon, and the collapse always results in a black hole.

As mentioned in the previous section, the condition $t_{ah}(r) > t_0$ is necessary, but not sufficient: local visibility requires the existence of an outgoing light-like geodesic emitting from the singularity with no trapped surfaces in its path. This can only happen if the apparent horizon forms \textit{sufficiently} late. In general, whenever the formation of the apparent horizon is sufficiently delayed, for example due to strong shearing effects \cite{Joshi2}, the singularity is exposed to external observers, at least locally, and becomes naked\footnote{In fact, all evidence indicates that the shear remains the crucial factor even when pressures are non-vanishing \cite{Joshi2}. In particular, the central singularity can become naked even when radial pressure remains non-negative throughout the cloud's evolution \cite{Cooperstock}.}. In black hole formation, instead, the apparent horizon forms before gravitational collapse and the singularity is hidden behind a global event horizon. 

Given that only the cases $n=1,2,3$ allow for naked singularities, we can check for local visibility by explicit construction. We assume an outgoing radial null geodesic starting at the singularity of the form \cite{Barve}
\begin{equation}\label{ansatz}
    t = t_0 + a \left( \frac{r}{r_0} \right)^{\alpha},
\end{equation}
to leading order in $r$, with $a,\alpha>0$. In order for the geodesic to lie in the ambient spacetime, we require $t \leq t_c(r)$, which by (\ref{tc}) is satisfied for all $\alpha > n$, and for $a<3 t_0/2(n+3)$ if $\alpha = n$.

To leading order in $r$, (\ref{solution}) is
\begin{equation}\label{Rexp}
    R = \left[ 1 - \left( 1-\frac{3}{2(n+3)} \left( \frac{r}{r_0}\right)^n\right)\frac{t}{t_0}\right]^{2/3}r.
\end{equation}
Differentiating with respect to $r$ we get
\begin{equation}\label{R'exp}
    R' = \left[ 1 - \left( 1-\frac{3}{2(n+3)} \left( \frac{r}{r_0}\right)^n\right)\frac{t}{t_0}\right]^{-1/3}\left[ 1 - \left( 1-\frac{2n+3}{2(n+3)} \left( \frac{r}{r_0}\right)^n\right)\frac{t}{t_0}\right].
\end{equation}
Given that by (\ref{ansatz}) $dt/dr = \alpha (a/r_0) (r/r_0)^{\alpha-1}$, (\ref{geodesic}) evaluated on the assumed geodesic gives
\begin{equation}\label{naked}
    \alpha \frac{a}{r_0} \left( \frac{r}{r_0}\right)^{\alpha-1} = \frac{1 - \left( 1-\frac{2n+3}{2(n+3)} \left( \frac{r}{r_0}\right)^n\right) \left( 1+ \frac{a}{t_0} \left( \frac{r}{r_0}\right)^\alpha\right)}{\left[ 1 - \left( 1-\frac{3}{2(n+3)} \left( \frac{r}{r_0}\right)^n\right)\left( 1+ \frac{a}{t_0} \left( \frac{r}{r_0}\right)^\alpha\right)\right]^{1/3}}
\end{equation}
If the equation above admits self-consistent solutions, the singularity is locally naked, meaning that there exists at least one outgoing null geodesics which terminates arbitrarily close to the singularity.

Let's consider first the case $\alpha > n$. At leading order, (\ref{naked}) gives
\begin{equation}
    \alpha \frac{a}{r_0} \left( \frac{r}{r_0}\right)^{\alpha-1} = \left( 1 + \frac{2n}{3} \right) \left( \frac{3}{2(n+3)}\right)^{2/3} \left(\frac{r}{r_0}\right)^{2n/3},
\end{equation}
which implies $\alpha = 1+2n/3$ and $a/r_0 = (3/2(n+3))^{2/3}$. The condition $\alpha > n$ translates to $n<3$, meaning that the singularity is locally naked for $n=1$ and $n=2$, confirming our previous analysis.

In the case $\alpha = n$, (\ref{naked}) gives
\begin{equation}
    n \frac{a}{r_0} \left( \frac{r}{r_0}\right)^{n-1} = \frac{\frac{2n+3}{2(n+3)} - \frac{a}{t_0}}{\left(\frac{3}{2(n+3)} - \frac{a}{t_0}\right)^{1/3}} \left( \frac{r}{r_0}\right)^{2n/3},
\end{equation}
which requires $n=3$. With this choice of $n$, the expression above reduces to
\begin{equation}\label{conditiona}
     3 y=  \frac{\frac{3}{4} - y}{\left( \frac{1}{4} - y\right)^{1/3}} \frac{r_0}{t_0},
\end{equation}
subject to the constraint $y < 1/4$, where $y \equiv a/t_0$. The equation above admits real solutions only when $t_0/r_0 > (4+2\sqrt{3})/3$, or, equivalently, when $r_0 > (28+16 \sqrt{3}) GM \approx 56 \, GM$, a stronger constraint than the one we deduced by just requiring $t_{ah} > t_0$.

\section{Global Visibility}\label{global}
In order for the singularity to be visible to observers at infinity, the geodesic is prohibited from crossing any apparent horizon throughout the collapsing cloud.

First, then, we need to check that there are no trapped surfaces on the initial hypersurface, i.e. we must require $r>2Gm$ for all $r \leq r_0$ at $t=0$. Given (\ref{mass}), the condition translates to
\begin{equation}\label{initialcond}
    \frac{2GM}{r_0} < \frac{n}{\left(n+3-3 x^{n}\right)x^2} \quad \, \text{for all } 0 \leq x\leq 1 ,
\end{equation}
with $x \equiv r/r_0$. The function on the right-hand side of the equation has a minimum at $x^n = (2n+6)/(3n+6)$, therefore (\ref{initialcond}) entails
\begin{equation}
    \frac{2GM}{r_0} < \frac{2+n}{3+n} \left(\frac{3n+6}{2n+6}\right)^{2/n}.
\end{equation}
The function on the right hand side is $\approx 1$ for $n=1,2,3$, so the constraint is merely the statement that the cloud is not a black hole initially.

The sufficient condition for global visibility was found in \cite{Jhingan} and is given by
\begin{equation}
    t_c'(r) > \frac{G}{3}(26+15 \sqrt{3}) \, m'(r) \quad \, \text{for all } 0\leq r \leq r_0.
\end{equation}
By (\ref{mass}) and (\ref{t_c}), this yields
\begin{equation}\label{condition}
    \frac{n}{\sqrt{2} (3+n)} \frac{x^{n-3}}{(1-x^n) \left( \frac{3+n}{n} - \frac{3}{n} x^n\right)^{3/2}} \left(\frac{r_0}{GM}\right)^{3/2} > 26+15 \sqrt{3} \quad \, \text{for all } 0\leq x\leq 1.
\end{equation}
Now, one has to distinguish the cases $n=1,2$ and $n=3$. In the former case, the function on the left hand side of (\ref{condition}) has a minimum at 
\begin{equation}
    x = \left(\frac{12+3n-\sqrt{25n^2+8n^3}}{6(2+n)}\right)^{1/n},
\end{equation}
therefore condition (\ref{condition}) is satisfied if
\begin{equation}
    \left(\frac{r_0}{GM} \right)^{3/2} > (26+15 \sqrt{3}) \frac{(3+n)(3+\sqrt{25+8n}) (7+2n+\sqrt{25+8n})^{3/2}}{2 \times 6^{\frac{3}{n}} (2+n)^{\frac{3(n+2)}{2n}}(12-n(\sqrt{25+8n}-3))^{\frac{n-3}{n}}}.
\end{equation}
This gives $r_0 \gtrsim 28 \, GM$ for $n=1$, and $r_0 \gtrsim 36 \, GM$ for $n=2$. In the latter case, the function on the left hand side of (\ref{condition}) has a minimum at $x=0$, therefore the condition becomes $r_0 > (28+16 \sqrt{3}) GM \approx 56 \, GM$. This is the same condition we obtained in the previous section where we studied local visibility. This means that in the special case $n=3$, if the singularity is locally naked, it is also globally naked.

To summarize, in marginally bound spherically symmetric collapse models, the central singularity is 
\begin{itemize}
    \item locally naked for $n=1,2$ and for $n=3$ if $r_0 \gtrsim 56 \, GM$,
    \item globally naked for $n=1$ if $r_0 \gtrsim 28 \, GM$, $n=2$ if $r_0 \gtrsim 36 \, GM$, and $n=3$ if $r_0 \gtrsim 56 \, GM$.
\end{itemize}
In all other cases, the singularity is hidden.

It is worth pointing out that, as shown in \cite{Joshi4, Joshi5}, naked singularities in LTB spacetimes are generic, in the sense that given an initial density profile for the cloud, there is a non-zero measure set of configurations leading to the formation of a naked singularity. Moreover, the choice of considering only radial geodesics to characterize the naked singularity is not overly restrictive, as it can be shown that the existence of future-directed non-radial null geodesics emanating from the singularity is guaranteed by the existence of the corresponding future-directed radial null geodesics \cite{Mena}.

\section{Absorption Efficiency}\label{absorption}
Having established that, given certain generic initial conditions, the singularity can be globally naked and the radiation emitted visible to an outside observer, we now turn to proving explicitly the detector's efficiency.

Assuming the source is kept stationary at a distance $\epsilon r_0$ from the singularity\footnote{The source can be kept in stationary orbit around the singularity precisely because the singularity is naked and there are no trapped surfaces around it \cite{Babar}.}, we can fire gravitons to infinity by suitably choosing the initial mass density. Will those gravitons also be absorbed with high probability? To determine that, we need to compute the optical depth of the graviton through the dust cloud, which is defined as
\begin{equation}
    \tau = \int \sigma n(t,r(t)) dt = \frac{G}{\mu} \int_{\epsilon r_0}^{r_0} \rho(t(r),r) R' dr,
\end{equation}
where $\sigma \sim G$ is the graviton absorption cross section and $n \equiv \rho/\mu$ the number density of atoms, with $\rho=m'/4 \pi R^2 R'$ and $\mu$ the mass of a single atom. The atoms have to be contained inside the detector, so their Compton wavelength should at least be smaller than the detector's radius, i.e. $\mu r_0 > 1$.

In terms of $x \equiv r/r_0$, and using (\ref{solution}) and (\ref{mass}), the integral is
\begin{equation}
    \tau = \frac{G}{\mu} \int_{\epsilon r_0}^{r_0} \frac{m'}{4 \pi R^2} dr = \frac{G \rho_0 r_0}{\mu}\int_\epsilon^1 \frac{1-x^n}{\left( 1-(1+a x^\alpha)\sqrt{1-\frac{3}{3+n}x^n}\right)^{4/3}} dx.
\end{equation} 
Since $ax^\alpha > 0$, the optical depth is bounded from below by
\begin{equation}\label{tau>}
    \tau >  \frac{G \rho_0 r_0}{\mu}\int_\epsilon^1 \frac{1-x^n}{\left( 1-\sqrt{1-\frac{3}{3+n}x^n}\right)^{4/3}} dx.
\end{equation}
The integral above is dominated by values around $x=0$, therefore it is well approximated by
\begin{equation}
\begin{split}
    \tau &> \frac{G \rho_0 r_0}{\mu}\int_\epsilon^1 \frac{1-x^n}{\left( 1-\sqrt{1-\frac{3}{3+n}x^n}\right)^{4/3}} dx \approx  \frac{G \rho_0 r_0}{\mu} \left(\frac{2(3+n)}{3}\right)^{4/3} \int_\epsilon^1 (1-x^n) x^{-\frac{4n}{3}} dx \\
    &=  \frac{G \rho_0 r_0}{\mu} \left(\frac{2(3+n)}{3}\right)^{4/3} \left( \frac{3\left( \epsilon^{1-\frac{4n}{3}} - 1\right)}{4n-3} + \frac{3\left( \epsilon^{1-\frac{n}{3}} - 1\right)}{3-n} \right).
\end{split}
\end{equation}
For all values $n \geq 1$, the optical depth can be made arbitrarily large in the limit $\epsilon \rightarrow 0$. This means that in models of spherically symmetric gravitational collapse with a mass density of the form (\ref{rho}) and $n=1,2,3$, if the initial size of the cloud is large enough graviton detection is both efficient and observable.

A particular caveat of this construction is that by taking the limit $\epsilon \rightarrow 0$ one gets arbitrarily close to the singularity where unknown quantum gravity effects can spoil the experiment. It's worth pointing out then that the limit $\epsilon \rightarrow 0$ is not strictly necessary, as all we want for efficient graviton detection is for the optical depth to be greater than one. This can be achieved for reasonable values of $\epsilon$ by tuning the initial parameters $r_0$, $M$ and $\mu$. For example, if one takes $n=2$, $\mu r_0 \sim 10^2$, and $r_0/GM \sim 10^2$, numerical integration of (\ref{tau>}) shows that the optical depth becomes order one for $\epsilon \approx 5 \times 10^{-3}$, meaning that the gravitational source is sitting at a distance $\epsilon r_0 = 0.5 GM$ from the singularity. Clearly, if $M$ is large enough, the source is undisturbed by putative quantum gravity effects. 

In a realistic scenario, one could consider a cloud of hydrogen atoms with $\mu \sim 5 \times 10^{-20} m_p$, and total mass $M=10 M_\odot \sim 10^{39} m_p$. Then, assuming $n=3$ and $r_0/GM \sim 10^2$, $\mu r_0 \sim 10^{22}$ and $\epsilon \sim 10^{-8}$, meaning that the source is at a distance of $\epsilon r_0 \sim 1 \, \text{cm}$ from the singularity. The Kretschmann scalar (\ref{kert}) at that distance is of order $\mathcal{K} = 48 G^2 m^2/R^6 \sim G^2M^2/\epsilon^{4n} r_0^6 \sim 10^{-72} m_p^4$, well below the Planck density\footnote{At very high densities, other quantum effects might come into play which rearrange the state of matter. However, atoms are not strictly necessary for the experiment to succeed: all we care about is that the final system has a mass gap.}.

\section{Conclusions}\label{conclusions}
We have shown, in a specific model of inhomogeneous gravitational collapse, that graviton detection can be both visible and efficient in the absence of an event horizon. This demonstrates that, if the CCC is violated, gravitons can be (efficiently) detected. The converse implication, namely that if gravitons can be detected then the CCC is necessarily violated, if true, is presumably much more difficult to prove. Thus, even though it is tempting to conclude that the CCC is the fundamental reason for the infeasibility of graviton detection, more work is needed to establish this connection\footnote{It is worth noting that there are other proposals in the literature on how quantum gravity could show itself in experiments, apart from direct graviton detection, such as indirect detection through quantum entanglement \cite{Marletto, Kanno} and other various tabletop experiments \cite{Carney3}. Our analysis here is concerned with \textit{efficient} and \textit{direct} graviton detection.}. 

However, there is some circumstantial evidence:

\begin{itemize}
    \item Gravitational wave detectors based on laser interferometry, such as LIGO, need a sensitivity of less than a Planck length to be able to detect single gravitons \cite{Dyson}. Resolving such distances is deemed impossible due to black hole formation \cite{Calmet}.
    \item Many thought experiments trying to establish the quantization of gravity \cite{Eppley, Page, Mari, Belenchia, Danielson} in the spirit of Bohr and Rosenfeld \cite{Bohr} fail to do so due to the Planck length acting as a fundamental limit on spatial resolution \cite{Mattingly, Baym, Rydving}, as above.
    \item The Gertsenshtein effect could in principle be used to detect single gravitons \cite{Dyson, Rothman}. However, it can be shown that such a detector is always inefficient at sub-horizon scales due to nonlinear electromagnetic effects that break quantum coherence \cite{Pal}. The cosmological event horizon effectively hides graviton-photon oscillations \cite{Anninos}, preventing single graviton detection.
    \item  Measurement of tensor modes in the CMB could be used to establish the quantization of gravity \cite{Krauss}. However, it can be shown that tensor modes are unobservable in models that hinder the growth of trans-Planckian fluctuations \cite{Bedroya, Bedroya2}. In these models, trans-Planckian fluctuations are hidden behind the cosmological horizon and thus unable to turn classical and affect macroscopic observations. 
\end{itemize}
Note that in the last two experiments listed, it is the cosmic event horizon that prevents their successful completion, and as such the CCC, as commonly understood, does not apply. However, one can strengthen the formulation of the CCC to include these cosmological cases as well. The CCC posits that any singularity in spacetime must lie behind an event horizon. This can be taken as a particular instance of a more general conjecture on the unobservability of quantum gravity effects in the universe, namely that Nature conspires to hide the quantization of the gravitational field behind event horizons, either astrophysical or cosmic. In this more general sense, the CCC prevents the successful completion of all graviton detection experiments listed above.

Finally, one could establish the quantization of the gravitational field by detecting its quantum-induced noise in the lengths of the arms of a LIGO-like gravitational wave detector \cite{Parikh, Parikh2}. This noise is negligible for coherent states but is greatly enhanced in thermal and squeezed states. However, as shown in \cite{Carney}, the simple observation of enhanced noise in the measuring apparatus cannot be used to distinguish a quantum model of the gravitational field from a classical one. In order to demonstrate quantization one would need to observe sub-vacuum levels of noise. Crucially, any deviation from noise at the standard quantum limit is proportional to the detector's efficiency \cite{Carney2}, which means that, even in the case of highly squeezed graviton states, the quantum signature of the gravitational field is unobservable \textit{unless} the detector is already highly efficient, and the evidence so far shows that efficient graviton detection always lies beyond an event horizon.

\end{document}